\documentclass[%
 reprint,
 superscriptaddress,
nofootinbib,
nobibnotes,
 amsmath,amssymb,
 aps,
 prb,
 dvipsnames, 
 floatfix
]{revtex4-2}

\usepackage{cmap}
\usepackage{graphicx}% Include figure files
\usepackage{dcolumn}% Align table columns on decimal point
\usepackage{bm}% bold math
\usepackage{xcolor}
\usepackage[hyperfootnotes=false,breaklinks=true]{hyperref}% add hypertext capabilities
\hypersetup{%
bookmarksopen=true,
bookmarksopenlevel=1,
colorlinks=true,   
linkcolor=blue,
anchorcolor=blue,
citecolor=blue,
filecolor=blue,
urlcolor=blue,
pdfpagemode=UseOutlines,
pdfstartview={XYZ null null 1.25},
linktocpage=true,
}

\newcommand{\eph}{e-ph}
\newcommand{\pdag}{{\phantom\dagger}}

\begin{document}

\title{Enhanced superconductivity in FeSe/SrTiO$_3$ from the combination of forward scattering phonons and spin fluctuations}

\author{Louk Rademaker}
\affiliation{Department of Theoretical Physics, University of Geneva, 1211 Geneva, Switzerland}

\author{Gustavo Alvarez-Suchini} 
\affiliation{Department of Physics and Astronomy, The University of Tennessee, Knoxville, Tennessee 37996, USA}

\author{Ken Nakatsukasa}
\affiliation{Department of Physics and Astronomy, The University of Tennessee, Knoxville, Tennessee 37996, USA}

\author{Yan Wang}
\affiliation{Quantum Computational Science Group, Oak Ridge National Laboratory, Oak Ridge, Tennessee 37831, USA}

\author{Steven Johnston}
\affiliation{Department of Physics and Astronomy, The University of Tennessee, Knoxville, Tennessee 37996, USA}

\date{\today}

\begin{abstract}
We study the effect of combining spin fluctuations and forward scattering electron-phonon ({\eph}) coupling on the superconductivity in the FeSe/SrTiO$_3$ system modeled by a phenomenological two-band Hubbard model with long-range {\eph} interactions. We treat the electron and phonon degrees of freedom on an equal footing using a \emph{fully} self-consistent FLEX plus Migdal-Eliashberg calculation, which includes a self-consistent determination of the spin fluctuation spectrum. Based on FeSe monolayers, we focus on the case where one of the bands lies below the Fermi level (i.e. incipient), and demonstrate that the combined interactions can enhance or suppress $T_c$, depending on their relative strength. For a suitable choice of parameters, the spin-fluctuation mechanism yields a $T_c \approx 46.8$ K incipient $s_\pm$ superconductor, consistent with surface-doped FeSe thin films. A forward-focused {\eph} interaction further enhances the $T_c$, as observed in monolayer FeSe on SrTiO$_3$.
\end{abstract}

\maketitle

\section{Introduction}

Monolayers of FeSe grown on oxide substrates like SrTiO$_3$ (STO) become superconducting at temperatures as high as $55-75$ K~\cite{Liu:2012di,He:2013cna,Wang:2012jx}, far in excess of the critical temperature of bulk FeSe ($T_c\approx 8$ K at ambient pressure~\cite{Hsu14262}). This discovery holds the promise of optimizing $T_c$ through heterostructure engineering \cite{LeeReview, CohHeterostructure, PhysRevLett.119.107003} once the microscopic mechanism(s) behind this phenomenon are identified and ultimately understood. Despite years of intense research, however, both the pairing mechanism and the gap symmetry of the FeSe monolayers remain unresolved \cite{Kreisel:2020ec}. 

Some aspects of the FeSe/STO system are now firmly established. The first aspect pertains to the role of doping. Superconducting FeSe monolayers are heavily electron-doped and their Fermi surface consists of only electron pockets centered at $(\pi/a,\pi/a)$ (the $M$-point) in the two Fe/unit cell notation \cite{He:2013cna}. In contrast, bulk FeSe is similar to other Fe-based superconductors \cite{Richard_2011, VANROEKEGHEM2016140} and has both electron pockets at $M$ and hole pockets at $\Gamma$. What is unclear is how much this doping contributes to the enhanced superconductivity. Empirically, electron doping seems to play a role, as evidenced by the fact that surface doping FeSe thin films pushes the hole-like band below the Fermi level ($E_\mathrm{F}$) and raises the $T_c$ to about 40 K~\cite{Miyata:2015kb}. On the one hand, the lack of hole-like bands crossing $E_\mathrm{F}$ speaks against Fermi-surface nesting arguments \cite{Mazin2008}. On the other hand, several studies have demonstrated that bands below $E_\mathrm{F}$ can still contribute to pairing \cite{Bang2014, Chen2015, Linscheid2016, Mishra2016}, provided the bands in question are not too far below $E_\mathrm{F}$, i.e. ``incipient" bands. 

For the FeSe monolayer, the incipient band pairing scenario implies that antiferromagnetic spin fluctuations contribute to pairing. Recently, a resonant inelastic X-ray scattering (RIXS) \cite{Pelliciari:2020tu} study succeeded in measuring and contrasting the spin excitations in FeSe in going from bulk to monolayer form. There, the authors reported a significant hardening and reorganization of the spin excitations in the monolayer, in agreement with QMC calculations \cite{Pelliciari:2020tu} of the bilayer Hubbard model~\cite{Maier:2011gp}. This observation reveals nontrivial changes in the magnetic excitation spectrum of the monolayer, which must have some impact on the resulting $T_c$. 

\begin{figure}[t]
    \centering
    \includegraphics[width=\columnwidth]{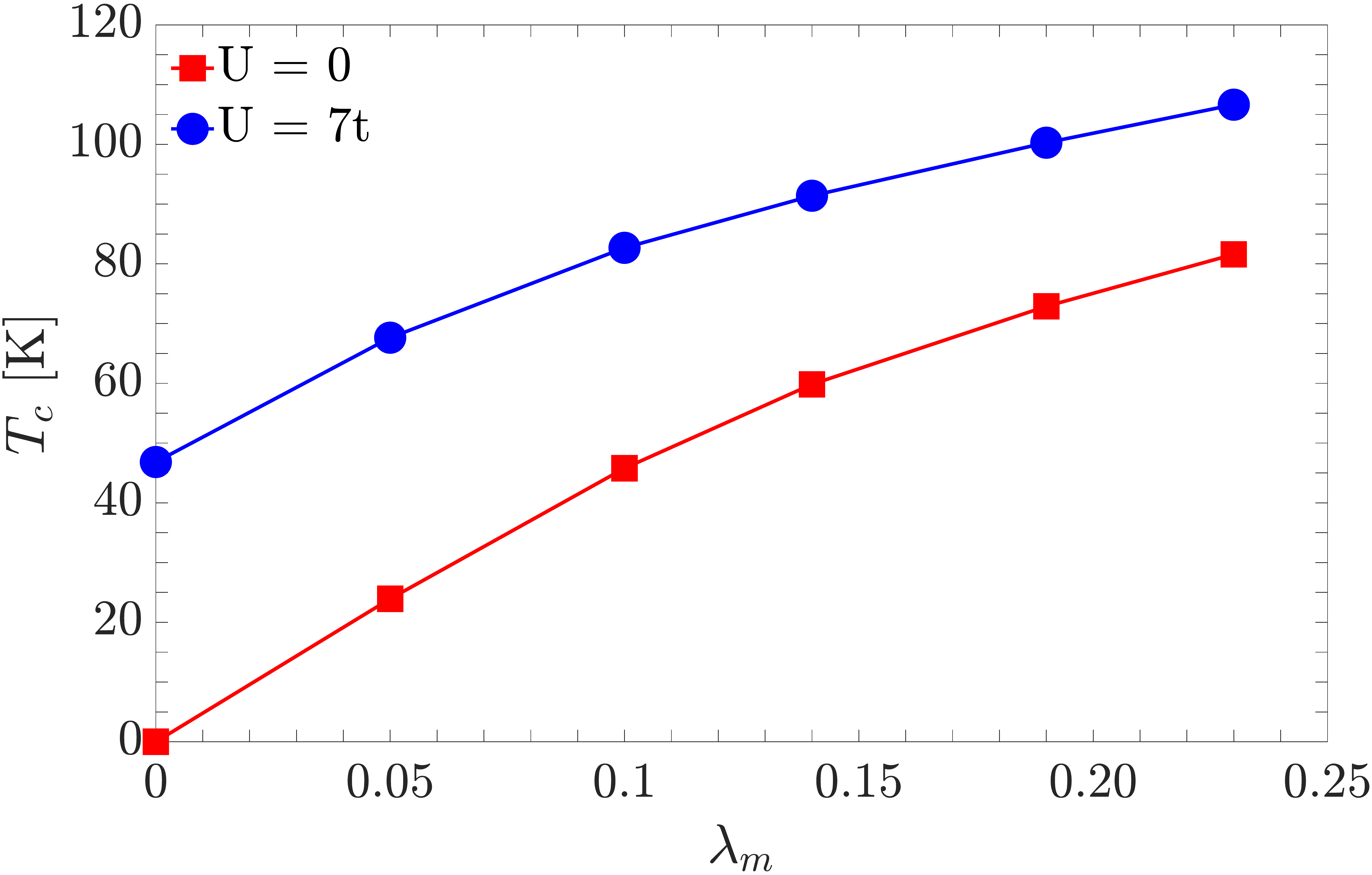}
    \caption{
    In the presence of pairing mediated by spin-fluctuations only, our model calculations ($U=7t$, $t = 75$ meV) for a monolayer of FeSe predict an $s_\pm$ incipient superconductivity with $T_c \approx 46.8$ K. The 
    addition of an {\eph} forward scattering further increases $T_c$ approximately linearly with $\Delta T_c \sim \lambda_m \Omega$, where $\lambda_m$ characterizes the strength of the {\eph} scattering and $\Omega$ the energy of the optical phonon mode (see Sec.~\ref{Sec:Model}). The slope of $\Delta T_c$ versus $\lambda_m$ is smaller for $U=7t$ (in blue) 
    compared to the $U=0$ case (in red), which is due to the quasiparticle renormalization induced by the Hubbard interaction. The values of $t_\perp$ are $4.21t$ and $2.36t$ for $U = 0$ and $7t$, respectively.
    } 
    \label{fig:FigTcvslambda}
\end{figure}

The second aspect of the FeSe/STO system that is now established is the role of the substrate, which provides an additional source of pairing. The existence of this contribution is evidenced by the fact that FeSe on STO realizes $T_c$ values that are consistently $15-25$ K higher than those achieved in FeSe intercalates with the same nominal doping but without the substrate. The microscopic mechanism behind this contribution remains unresolved, however. The proposal that interests us here 
is that there is a cross-interfacial {\eph} coupling between the FeSe electrons and optical oxygen phonon modes in the substrate. This scenario gained traction following the observation of replica bands in the electronic structure probed by angle-resolved photo-emissions spectroscopy (ARPES) \cite{Lee2014}, which can be explained by an {\eph} interaction that is peaked sharply around zero momentum transfer ${\bf q} = 0$. Such couplings are of great interest because they can produce large values of $T_c$ when treated within the Migdal-Eliashberg (ME) formalism~\cite{PhysRevB.49.4395, Rademaker2016, WangSUST, Kulic2017}. 
There are several challenges to this hypothesis, however. Density functional theory (DFT)~\cite{Wang2016} and experiments on the phonon linewidth~\cite{Wang2017, Zhang2018} suggest the coupling is weaker than the value needed to reproduce the strength of the replica bands. There also exists criticism that forward scattering will be ineffectual in the presence of Coulomb interaction~\cite{ZhouMillis1, ZhouMillis2}. Moreover, quantum Monte Carlo (QMC) on small clusters calculations don't reproduce the large $T_c$'s obtained by ME theory \cite{XiXiang2019}, which suggests that vertex corrections may be significant. However, a recent study using an analytic approach found that $T_c$ is {\it enhanced} with the inclusion of the vertex corrections \cite{LiuPreprint}. The source of this discrepancy is unknown at this time. Some have also argued that the replica bands are not related to an intrinsic {\eph} coupling but rather an interaction of the outgoing photo-electron with the substrate phonons~\cite{PhysRevLett.120.237001}.

This situation has lead to proposals where the {\eph} coupling boosts the $T_c$ established by another unconventional mechanism~\cite{Lee2014, LI2016925, XiXiang2019, Bang2019}. This idea hinges on the forward-focused nature of the {\eph} interaction, which can mediate attractive interactions in all pairing channels \cite{VARELOGIANNIS20071125}. Two recent experimental studies have bolstered this scenario. One was an ARPES experiment that established a linear correlation between the strength of the {\eph} coupling and the size of the superconducting gap in FeSe monolayers on STO~\cite{Song:2019cg}. The second was a comparative study of correlated monolayers of FeSe and uncorrelated monolayers of FeS, both grown on STO \cite{Shigekawa24470}. There, the authors resolved replica bands in both systems but found that only the FeSe/STO interface exhibited a superconducting state. 

Based on this evidence, it is quite likely that both spin fluctuations and {\eph} coupling are relevant to the FeSe monolayers. Until recently \cite{Schrodi:2020bq}, however, no studies have been carried out for this system where these two contributions are treated on an equal footing. Here, we address this question directly and study the effects of combined spin fluctuations and small-${\bf q}$ {\eph} coupling using a fully self-consistent ME + Fluctuation-Exchange (FLEX) formalism. To model the spin fluctuations, we use the two-band (bilayer) Hubbard model. While this model does not capture all five Fe $d$-bands, it captures the electron-pocket phenomenologically at the \textit{M}-point and an incipient hole-pocket at $\Gamma$. Notably, the simplicity of the model allows us to calculate the ME+FLEX equations fully self-consistently, where the electron self-energy is fed back into the interaction kernels at each step. 
This aspect of our approach differs from the related work in Ref.~\cite{Schrodi:2020bq}, and we argue it is crucial for unifying the incipient $s_\pm$ and forward-scattering {\eph} scenarios described above. 

Our main result is summarized in Fig.~\ref{fig:FigTcvslambda}. By adopting a reasonably strong Hubbard repulsion, we can obtain an $s_\pm$ superconducting solution with a 
$T_c = 46.8$ K. The inclusion of a forward-focused {\eph} coupling leads to a further quasi-{\em linear} enhancement of the $T_c$, consistent with experiments. We find that the enhancement is not as dramatic as in the previously studied case of only forward scattering~\cite{Rademaker2016, WangSUST} due to the increase in the quasiparticle mass introduced by  a non-zero Hubbard $U$. Nonetheless, we can understand the enhancement of $T_c$ using a simple model of two pairing channels working in parallel. Our result also matches qualitatively the experimental correlation between a larger spectral gap and the spectral weight of the replica bands in FeSe/STO~\cite{Song:2019cg}. (The spectral weight of the replica band is directly proportional to $\lambda_m$, see Ref.~\cite{Rademaker2016}.) 
Our results, therefore, demonstrate the feasibility of combining incipient-band pairing with cross-interfacial {\eph} coupling to produce high-$T_c$ superconductivity. 

The organization of this work is as follows. Sections \ref{Sec:Model}A \& \ref{Sec:Model}B present our model and discusses our choice of parameters in relation to the FeSe/STO system. 
Next, Sec.~\ref{Sec:Model}C discusses the details of our combined FLEX+ME approach for treating the spin-fluctuations and forward scattering {\eph} interactions, while Sec.~\ref{Sec:Model}D provides additional computational details. We present our main results in Sec.~\ref{Sec:Results}. Finally, Sec.~\ref{Sec:Outlook} summarizes our conclusions and provides some outlook for future work. 
%%%%%%%%%%%%%%%%%%%%%%%%%%%%%%%%%%%%%%%%%%%%%%%%%%%%%%%%%%%%%%%%
\section{Model \& Methods}
\label{Sec:Model}

\subsection{Model}
We model the FeSe monolayer system using a two-band Hubbard model \cite{Mishra2016,Linscheid2016,Pelliciari:2020tu} defined on a square lattice with additional {\eph} interactions. The model Hamiltonian is partitioned as $\hat{H} = \hat{H}_{\text{el}} + \hat{H}_{\text{ph}} + \hat{H}_{\text{\eph}}$. The Hamiltonian of the electronic system is given by  
\begin{align}
    \hat{H}_{\text{el}} &= -t\sum_{\langle i,j\rangle,\alpha, \sigma} 
    c^\dagger_{i,\alpha,\sigma}c^\pdag_{j,\alpha,\sigma} 
    - t^\pdag_\perp \sum_{i,\sigma} \left[c^\dagger_{i, 1, \sigma} c^\pdag_{i, 2, \sigma} + \mathrm{h.c.}\right] \nonumber \\ &
    - \mu \sum_{i, \alpha, \sigma} n_{i, \alpha, \sigma}
    + U \sum_{i \alpha} n_{i , \alpha, \uparrow} n_{i, \alpha, \downarrow}, \label{Eq:Hel}
\end{align}
where $c^\dagger_{i,\alpha,\sigma}$ ($c^\pdag_{i,\alpha,\sigma}$) creates (annihilates) an spin $\sigma$ ($=\uparrow,\downarrow$) electron in orbital (or layer) $\alpha$ ($=1,2$) of unit cell $i$; $\langle \dots \rangle$ denotes a sum over nearest neighbors; $U$ is the on-site Hubbard repulsion; and, $t>0$ and $t_\perp \geq 0$ are the intra- and interlayer hopping integrals, respectively. 

\begin{figure}[t]
    \centering
    \includegraphics[width=\columnwidth]{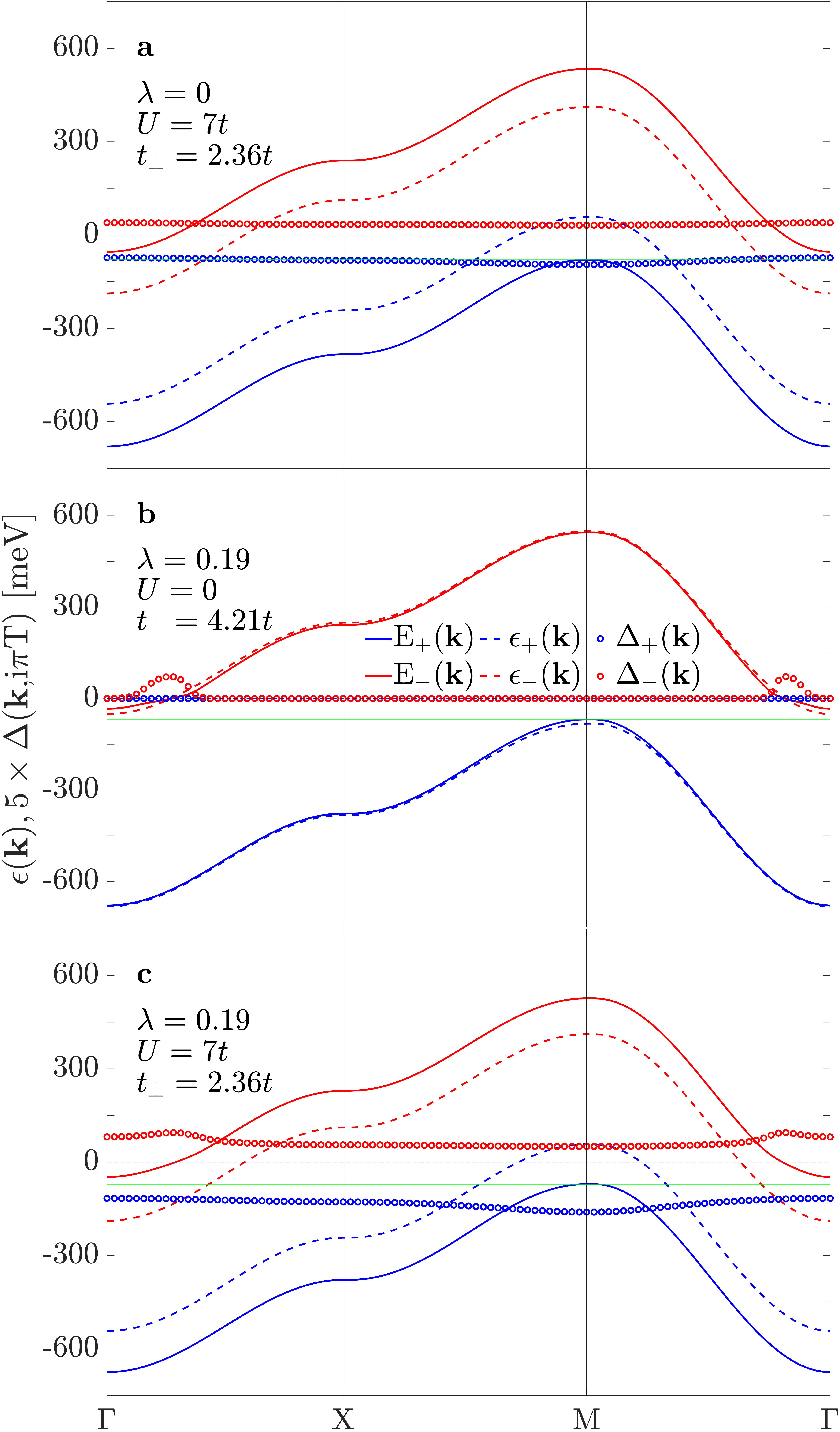}\vspace{-0.1cm}
    \caption{The bare [$\epsilon_\pm({\bf k})$, dashed lines] and renormalized 
    [$E_\pm({\bf k})$, solid lines] band structure and superconducting gap 
    $\Delta_\pm({\bf k})$ [open dots] in our model along 
    the high symmetry cuts of the first Brillouin zone. The parameters  
    $\{t_\perp, \lambda_m, U\}$ for the top, middle, and bottom 
    panels are $\{2.36t,~0,~7t\}$, $\{4.21t,~0.19,~0\}$, and 
    $\{2.36t,~0.19,~7t\}$, respectively. 
    The values of $t_\perp$ were selected to 
    position the top of the \textit{renormalized} hole-like band at 
    $M$ approximately $60$ to $80$ meV below the Fermi level (as indicated by the 
    thin green line). The chemical potential in each case is 
    adjusted during the self-consistency loop to produce a total 
    filling of $n = 2.1$~$e$/unit cell. Finally, note that $\Delta_+({\bf k})$ and $\Delta_-({\bf k})$ are degenerate in panel b.}
    \label{fig:Bands}
\end{figure}

When $U = 0$, Eq.~(\ref{Eq:Hel}) can be diagonalized exactly, 
resulting in two bands whose dispersions  
\begin{equation*}
    \epsilon_\pm({\bf k}) = -2t\left[\cos\left(k_xa\right)+\cos\left(k_y a\right)\right]-\mu\mp t_\perp
\end{equation*} 
are separated by $2t_\perp$. 
For intermediate values of the interlayer hopping $0 < t_\perp < 2t$ and near the electron-hole compensated filling $n=2$, the bare dispersion consists of a hole-like band centered at ${\bf k} = (\pi/a,\pi/a)$ (the $M$-point) and an electron-like band centered at ${\bf k} = (0,0)$ (the $\Gamma$-point), 
as shown in Fig.~\ref{fig:Bands}(a) by the dashed lines. (Note that the situation is 
reversed in FeSe/STO, where the hole- and electron-like bands are centered at the $\Gamma$- and $M$-points, respectively. Our band structure can be made equivalent to this by 
applying a 
particle-hole transformation.) Further increasing the value of $t_\perp$ for a fixed filling 
drives the hole-like band at $\Gamma$ to energies deeper below the Fermi energy, and for sufficiently large $t_\perp$, this band can be made incipient in the non-interacting case.  For this reason, the bilayer model has attracted considerable interest as a simple toy Hamiltonian that can describe systems with and without an incipient band by tuning a single parameter \cite{PhysRevB.66.184508,PhysRevB.72.212509,PhysRevB.77.144527,Maier:2011gp,Mishra2016,Matsumoto2020,PhysRevResearch.2.023156,PhysRevB.99.140504}. 

The terms $\hat{H}_{\text{ph}} + \hat{H}_{\text{\eph}}$ model the \emph{substrate}
lattice degrees of freedom and their interaction with the electrons. Here,   
\begin{equation}
    \hat{H}_{\mathrm{ph}} = \Omega \sum_{{\bf q}} \left(b^\dagger_{{\bf q} }b^{\phantom\dagger}_{\bf q}+\tfrac{1}{2}\right)
\end{equation}
describes a non-interacting dispersionless transverse optical phonon branch;  
$b^\dagger_{\bf q}$ ($b^{\pdag}_{\bf q}$) creates (annihilates) a phonon mode with momentum ${\bf q}$; 
and $\Omega = 100$ meV is the energy of the mode. Finally, we model the coupling between the phonons and the electronic subsystem using a long-range {\eph} interaction 
\begin{equation}
     \hat{H}_{\text{\eph}} = 
        \sum_{\substack{{\bf k}, {\bf q}\\\alpha, \sigma}} g({\bf q})
        c^\dagger_{{\bf k}+ {\bf q},\alpha,\sigma}
        c^{\phantom\dagger}_{{\bf k},\alpha,\sigma} 
        (b^\dagger_{-{\bf q}} + b^{\phantom\dagger}_{\bf q}), 
\end{equation}
where $g({\bf q})$ is the matrix element of the {\eph} interaction. 
Here, we adopt $g({\bf q}) = g_0 e^{- |{\bf q}|/q_0}$, which is appropriate 
for the proposed forward-focused cross-interfacial {\eph} interaction that is believed 
to be relevant across the FeSe/oxide interface~\cite{Lee2014,Rademaker2016,WangSUST}. 

\subsection{Model Parameters}
The parameters $t$, $t_\perp$, and $\mu$ are selected to reproduce the qualitative features of monolayer FeSe. Specifically, we fix the in-plane hopping parameter to $t = 75$ meV to facilitate comparisons with our previous work~\cite{Rademaker2016,WangSUST}, while the value of $\mu$ is adjusted to maintain a fixed filling 
of $n = 2.1$ $e$/unit cell. These choices result in a renormalized electron pocket crossing $E_\mathrm{F}$ whose size and shape are similar to the band observed in FeSe monolayers~\cite{Rademaker2016}. We also adjust the value of $U$ to control the strength of the electronic correlations. 
When $U\ne 0$, the Hartree contribution to the self-energy produces 
an additional shift of the renormalized bands. To account for this, we adjusted the value of $t_\perp$ to ensure that the top of the incipient hole-like band remains $\approx 60$ to $80$ meV below the Fermi level, consistent with experiments~\cite{He:2013cna, Lee2014, Shigekawa24470}. The effect of this Hartree shift can be seen in Fig.~\ref{fig:Bands}, and can be quite sizeable: for large $U$ the {\em bare} bandstructure (without Hartree contribution) contains electron {\em and} hole-like bands while the \emph{renormalized} hole-like band is completely below the Fermi level.

Previous comparisons to ARPES data ~\cite{Lee2014,Rademaker2016} suggest that 
the forward scattering range (in momentum space) is very short 
with $q_0a = 0.1 - 0.3$, where $a$ is the in-plane lattice constant of the FeSe monolayers. Here, we adopt $q_0a = 0.1$, in line with our prior work~\cite{Rademaker2016}. 
The strength of the {\eph} coupling $g_0$ is adjusted in the absence of the Coulomb interaction to yield a given Fermi surface 
average mass enhancement $\lambda_m$, which is calculated using a 
Fermi surface average of the mass renormalization 
\begin{equation*}
    \lambda_m = \left\langle -\frac{\partial \mathrm{Re}\Sigma({\bf k},\omega)}{\partial \omega}\bigg|_{\omega=0} \right\rangle_\text{FS}
    \approx \left\langle -\frac{\mathrm{Im}\Sigma({\bf k},\text{i}\pi T)}{\pi T} \right\rangle_\text{FS}.
\end{equation*}
The Fermi surface average is calculated with $\left\langle f(\mathbf{k}) \right\rangle_\text{FS} = \sum_{\mathbf{k}} f(\mathbf{k}) \delta(\epsilon_{+}(\mathbf{k})) / \sum_{\mathbf{k}} \delta(\epsilon_{+}(\mathbf{k}))$.
Note that $\lambda_m$ depends on the temperature, the phonon energy $\Omega$, and 
state (normal vs superconducting) of the system~\cite{WangSUST, Kulic2017}. We, therefore, adopt the convention that $\lambda_m$ is reported as its value at $T = 100$ K and in the normal state. 

\subsection{Self-Consistent ME+FLEX Formalism}
We treat the electronic and phonon-mediated interactions on the same level by combining the FLEX diagrams~\cite{Bickers:1989hu, Linscheid2016, Nocera2018} with the ME diagrams~\cite{Rademaker2016,WangSUST}. Our implementation solves the resulting equations self-consistently and accounts for both the electron self-energy and its effect on the pairing interactions while also retaining the full momentum dependence of both sets of quantities~\cite{Nocera2018}. This treatment is enabled by our efficient implementation of the FLEX and ME formalism, as outlined in Refs.~\cite{Nocera2018} and \cite{DeePRB2019}, respectively. 

Before proceeding, we would like to note the similarities and differences between our treatment of the problem and the one presented in Ref.~\cite{Schrodi:2020bq}.   
Both that work and ours treat the electronic and phononic degrees of freedom at the same diagrammatic level; however, Ref.~\cite{Schrodi:2020bq} solved the problem using a five-band model for FeSe that was derived from DFT calculations. Their model, therefore, includes all five Fe $3d$ orbitals and the corresponding multi-orbital Hubbard and Hund's interactions. But it also required ad hoc modifications of some non-interacting band parameters to account for lattice 
mismatch between the FeSe monolayer and the STO substrate. Our model uses the two-band Hubbard model with only intra-orbital Hubbard interactions to describe the electronic degrees of freedom and is, therefore, more phenomenological. The use of this simplified model, however, allows us to account for the feedback between the electron self-energies and the effective interaction by recomputing the latter with the dressed Green's functions during each stage of the self-consistency loop. This aspect is in contrast to Ref.~\cite{Schrodi:2020bq} (see also Ref.~\cite{Schrodi:2020bz}), which computed the effective interaction using the bare band structure and then held it fixed during the self-consistent calculation of the electron self-energies. This difference is important because the Hartree shift can result in self-consistently calculated effective interactions that are qualitatively different than the ones obtained using the bare band structure when $U$ is large. Both approaches have their merits and limitations, which we will discuss further in what follows.

The central objects to compute are the normal 
\begin{equation}
    G^{\sigma}_{ab} ({\bf k},\tau) =
        - \langle T^\pdag_\tau c^\pdag_{{\bf k},a,\sigma} (\tau) c^\dagger_{{\bf k},b,\sigma} (0) \rangle 
\end{equation}        
and anomalous         
\begin{equation}
    F^{\sigma \overline{\sigma}}_{ab}({\bf k},\tau) = 
         - \langle T^\pdag_\tau c^\pdag_{{\bf k},a,\sigma} (\tau)
         c^\pdag_{-{\bf k},b,\overline{\sigma}} (0)  \rangle.
\end{equation}
electron Greens functions in imaginary time in orbital space [$a$ and $b$ are 
orbital (layer) indices]. In the absence of any interactions ($U = \lambda_m = 0$), 
the bare Greens functions in the normal state and on the Matsubara frequency axis are  
$G_{0,\sigma}^{-1}({\bf k},\mathrm{i}\omega_n) = (i \omega_n -\mu)\mathbb{I} - \hat{H}_0({\bf k})$ and $F_0({\bf k},\mathrm{i}\omega_n)=0$, where $\hat{H}_0(\bf k)$ is the Fourier transform of the non-interacting Hamiltonian in orbital space and $\mathbb{I}$ is a $2\times 2$ identity matrix. As long as certain symmetries are preserved, as in the case of singlet pairing with $s$- and $d$-wave order parameters, all the matrix elements of the Nambu-Gorkov Green's functions can be determined from the $(\uparrow,\uparrow)$-block (normal Green's function $G_{\uparrow\uparrow}$, a $2\times 2$ matrix in orbital space) and the $(\uparrow,\downarrow)$-block (anomalous Green's function $F_{\uparrow\downarrow}$). Therefore, We will drop the spin index on $G$ and $F$ for brevity. 

The effect of interactions is captured by the normal and anomalous self-energy matrices, denoted $\Sigma_{ab} ({\bf k},i \omega_n)$ and $\Phi_{ab}({\bf k},i \omega_n)$, respectively. The dressed Green's functions are related to the bare ones and self-energies by using Dyson's equation for full Nambu-Gorkov Green's function, which can be simplified into two equations of matrices in orbital space
\begin{equation}
    G^{-1} = G_0^{-1} - \Sigma - \Phi [ - (G_0^*)^{-1} + \Sigma^* ]^{-1} \Phi^* 
\end{equation}
and 
\begin{equation}
    F^{-1} = - \left[ (G_0^*)^{-1} - \Sigma^* \right](G \Phi)^{-1} . 
\end{equation}
Here, $(^*)$ denotes complex conjugation, and all Greens functions and self-energies are written in orbital space and are understood to be functions of momentum ${\bf k}$ and fermionic Matsubara frequency $\omega_n = \pi T (2n+1)$, where $T$ is the temperature. 

In the FLEX formalism, we assume that the effective electronic interaction is mediated by spin and charge fluctuations, which are captured by spin and charge susceptibilities. Without inter-orbital Coulomb interactions, the effective interactions \cite{Schmalian1998,Kontani1998} only depend on the matrix elements of irreducible spin and charge susceptibility matrices that are given by 
\begin{equation}\label{eq:Chis0}
    \chi^{0,s}_{ab} (q) = - \frac{T}{N} \sum_k \left[ G_{ab} (k+q) G_{ba} (k)
        + F_{ab} (k+q) F^*_{ba}(k) \right] 
\end{equation}
and 
\begin{equation}\label{eq:Chic0}
    \chi^{0,c}_{ab} (q) = - \frac{T}{N} \sum_k \left[ G_{ab} (k+q) G_{ba} (k)
        - F_{ab} (k+q) F^*_{ba}(k) \right], 
\end{equation}
respectively. Here, we have adopted the 4-vector notation $k \equiv ({\bf k}, \text{i} \omega_n)$ and $q = ({\bf q}, \text{i} \omega_m)$, where $\omega_n = (2n+1) \pi T$ and $\omega_m = 2m \pi T$ are fermionic and bosonic Matsubara frequencies. 

The RPA spin and charge susceptibilities are
\begin{align}
    \chi^s &= ( \mathbb{I} - \chi^{0,s} U )^{-1} \chi^{0,s}, \\
    \chi^c &= ( \mathbb{I} + \chi^{0,c} U )^{-1} \chi^{0,c}.
\end{align}
The normal and anomalous FLEX interactions, which are also matrices in orbital space, are
\begin{equation}\label{Eq:FLEX_Int1}
    V^n (q) = \frac{3U^2}{2} \chi^s (q) + \frac{U^2}{2} \chi^c (q) - \frac{U^2}{2} (\chi^{0,s} + \chi^{0,c}) + U\mathbb{I}
\end{equation}
and
\begin{equation}\label{Eq:FLEX_Int2}
    V^a (q) = \frac{3U^2}{2} \chi^s (q) - \frac{U^2}{2} \chi^c (q) - \frac{U^2}{2} (\chi^{0,s} - \chi^{0,c}) + U\mathbb{I} .
\end{equation}

The above equations describe the effective interactions mediated by the spin and charge fluctuations in the system. The last constant term $U$ in each equation corresponds to Hartree-Fock contributions \cite{Bickers1989}. In addition to these, we also account for the {\eph} interaction at the level of ME theory. Because the interaction is assumed to be significantly peaked at small momentum transfers, we assume that the {\eph} interaction is diagonal in band space. This type of coupling results in an effective electron-electron interaction with equal intra- and inter-orbital components with
\begin{align}\nonumber
    V^{ep} (q)&= |g({\bf q})|^2 D_0(q) \left[\mathbb{I}+\sigma_x\right] \\
    &= -|g({\bf q})|^2 \frac{2 \Omega}{\Omega^2 + \omega_m^2} \left[\mathbb{I}+\sigma_x\right],\label{Eq:eph_Int}
\end{align}
where  $\sigma_x$ is the usual Pauli matrix. The Hartree contribution due to {\eph} interaction is assumed to be already included in the parameters of the bare band structure.

Combining the interactions from both sources then leads to self-consistent equations for the self-energies
\begin{align}
    \Sigma_{ab} (k) &= \frac{T}{N} \sum_q [ V_{ab}^n (q) - V^{ep}_{ab} (q) ] G^{\phantom a}_{ab} (k-q),
    \label{Eq:SelfEnergy1} \\
    \Phi_{ab} (k) &= \frac{T}{N} \sum_q [ V_{ab}^a (q) + V^{ep}_{ab} (q) ] F^{\phantom a}_{ab} (k-q).
    \label{Eq:SelfEnergy2} 
\end{align}
These equations are then solved iteratively until convergence is reached. Note that for Hartree-Fock term, a convergence factor $e^{i\omega_n 0^{+}}$ must be added for the Matsubara frequency sum.

We again emphasize that we recompute the spin and charge susceptibilities at every step, which allows both the phonon-mediated and spin- and charge-fluctuation-mediated interactions to talk to one another. It also ensures that the final spin and charge susceptibilities are obtained using the renormalized bandstructure rather than the bare bandstructure, which can be qualitatively different (see the discussion on the Hartree shift and Fig.~\ref{fig:Bands}). Another important consequence of this self-consistency procedure is that it allows us to reach the strong coupling regime. This regime includes interaction strengths that lay beyond the maximal value of $U$ imposed by the Stoner criterion (i.e. the divergence in the non-self-consistently computed susceptibility \cite{Schrodi:2020bz}). In fact, we find that for $U>5t$ the bare bandstructure produces divergent susceptibilities that are avoided in the final self-consistent solution.

To estimate the renormalized bandstructure of the system, we diagonalized the matrix
\begin{equation*}
    \left[\begin{array}{cc}\epsilon_0({\bf k})-\mu+\Sigma_{11}({\bf k},\mathrm{i}\pi T) & -t_\perp + \Sigma_{12} ({\bf k},\mathrm{i}\pi T)\\
    -t_\perp + \Sigma_{21} ({\bf k},\mathrm{i}\pi T) & \epsilon_0({\bf k})-\mu+\Sigma_{22}({\bf k},\mathrm{i}\pi T) \end{array}\right], 
\end{equation*}
where $\Sigma_{\alpha\beta}$ denotes the components of the normal self-energy in orbital 
space and $\epsilon_0({\bf k}) = -2t[\cos(k_xa)+\cos(k_ya)]$. The eigenstates of 
this matrix, with self-energy at lowest Matsubara frequency, are a good approximation to the dispersions inferred from the spectral function at low $T$ and low energy. They can, therefore, provide a good measure of the locations of the top of the hole-like band and the bottom of the electron-like band, which are located near $E_\mathrm{F}$. This approximation becomes less reliable far from the Fermi level. We label these approximate band dispersions as $E_\pm({\bf k})$ in Fig. \ref{fig:Bands}. 

\subsection{Computational details}\label{Appendix:ComputationalDetails}
We solved Eqs.~(\ref{eq:Chis0})-(\ref{Eq:SelfEnergy2}) self-consistently using a \mbox{MATLAB} code.\footnote{The version of the code used to generate this work and the 
corresponding data can be downloaded at  \url{https://github.com/JohnstonResearchGroup/Rademaker_etal_BilayerFLEX+ME_2021}} 
Our implementation uses fast Fourier Transforms to evaluate the necessary convolutions in momentum and Matsubara frequency space, as described in Ref.~\cite{DeePRB2019}. The momentum grid has $N_k = 4096$ momentum points, and for each temperature we used a Matsubara frequency cut-off at $\omega_{\mathrm{max}} = \tfrac{5}{2}W$, where $W$ is the total bandwidth of the non-interacting system. Beyond this cutoff, the terms in the Matsubara sum are approximated with their non-interacting values and summed to infinite frequencies by evaluating the product in the imaginary time domain.

For a given set of input parameters, we start our calculations at low temperature to find a stable symmetry-broken state iteratively using a superconducting ansatz. In practice, we initialize with a momentum-independent gap; nevertheless, small numerical roundoff errors can tip the solution towards an anisotropic (e.g. $d$-wave solution) if we iterate the self-consistency loop for long enough. As such, our code can find solutions in higher angular momentum channels (see for example Fig. 10 of Ref.~\cite{WangSUST}). 

Since we choose a constant energy cut-off, more Matsubara frequencies are included at lower temperatures, which increases the overall iteration time. But the number of iterations needed for convergence is much fewer in the superconducting state at low $T$ than near $T_c$, so we start our calculation from the lowest temperature point. Once the low-temperature solutions are obtained, we increase the temperature in small steps and initialize each new simulation with a self-energy obtained by interpolating on the Matsubara frequency grid solution from the previous temperature. 

When computing the charge and spin susceptibilities using the dressed Greens functions, we sometimes find that the susceptibility diverges before convergence is reached. This occurs because the Stoner criterion $\det \left(\mathbb{I}-U\chi^{0,s} \right)=0$ is met at some ${\bf q}$ for the current non-converged $\chi^{0,s}$. In those cases, we artificially cut-off the maximum in $[\mathbb{I}-U\chi^{0,s}]^{-1}$ to allow the simulation to continue iterating while we slowly remove the cut-off. By following this procedure, we can obtain final self-consistent solutions that are difficult to reach otherwise. We stress that our final results never have any divergences in the converged susceptibilities. 

Finally, as mentioned previously, the difference of Hartree-Fock terms between two bands can produce a sizable relative shift of the electron and hole bands. Depending on the coupling strengths $U$ and $\lambda_m$, we adjusted the bare interband coupling $t_\perp$ to ensure that the hole-like band was ca. $60-80$ meV below the Fermi level in the final self-consistent solutions. Similarly, during each step, we adjusted the chemical potential $\mu$ to ensure that we maintain a filling $n = 2.1$ $e$/unit cell.

%%%%%%%%%%%%%%%%%%%%%%%%%%%%%%%%%%%%%%%%%%%%%%%%%%%%%%%%
\section{Results}
\label{Sec:Results}
\subsection{Superconductivity without electron-phonon coupling}
We first consider the case without the {\eph} interaction ($U\neq0, \lambda_m =0$). Consistent with Refs.~\cite{Chen:2015dw,Linscheid2016}, we find a superconducting solution with an {\em incipient} $s_\pm$ order parameter for $U/t > 5$. For example, Fig.~\ref{fig:Bands}a plots the renormalized bands and superconducting gap along the high-symmetry cuts of the Brillouin zone for $t_\perp = 2.36t$ and $U = 7t$. In the non-interacting limit, both of the bare electronic bands cross the Fermi level, as indicated by the dashed blue and red lines. Once $U\ne 0$, the interband Hartree contribution to the self-energy further splits the bands and renders hole-like band incipient in the converged solution. We also find that the resulting gap function $\Delta_{\pm}({\bf k})$, where $\pm$ refers to the bands $\epsilon_\pm({\bf k})$, is weakly momentum dependent throughout the zone, and changes sign between the two bands, characteristic of an $s_\pm$ pairing symmetry. 

\begin{figure}[t]
    \centering
    \includegraphics[width=\columnwidth]{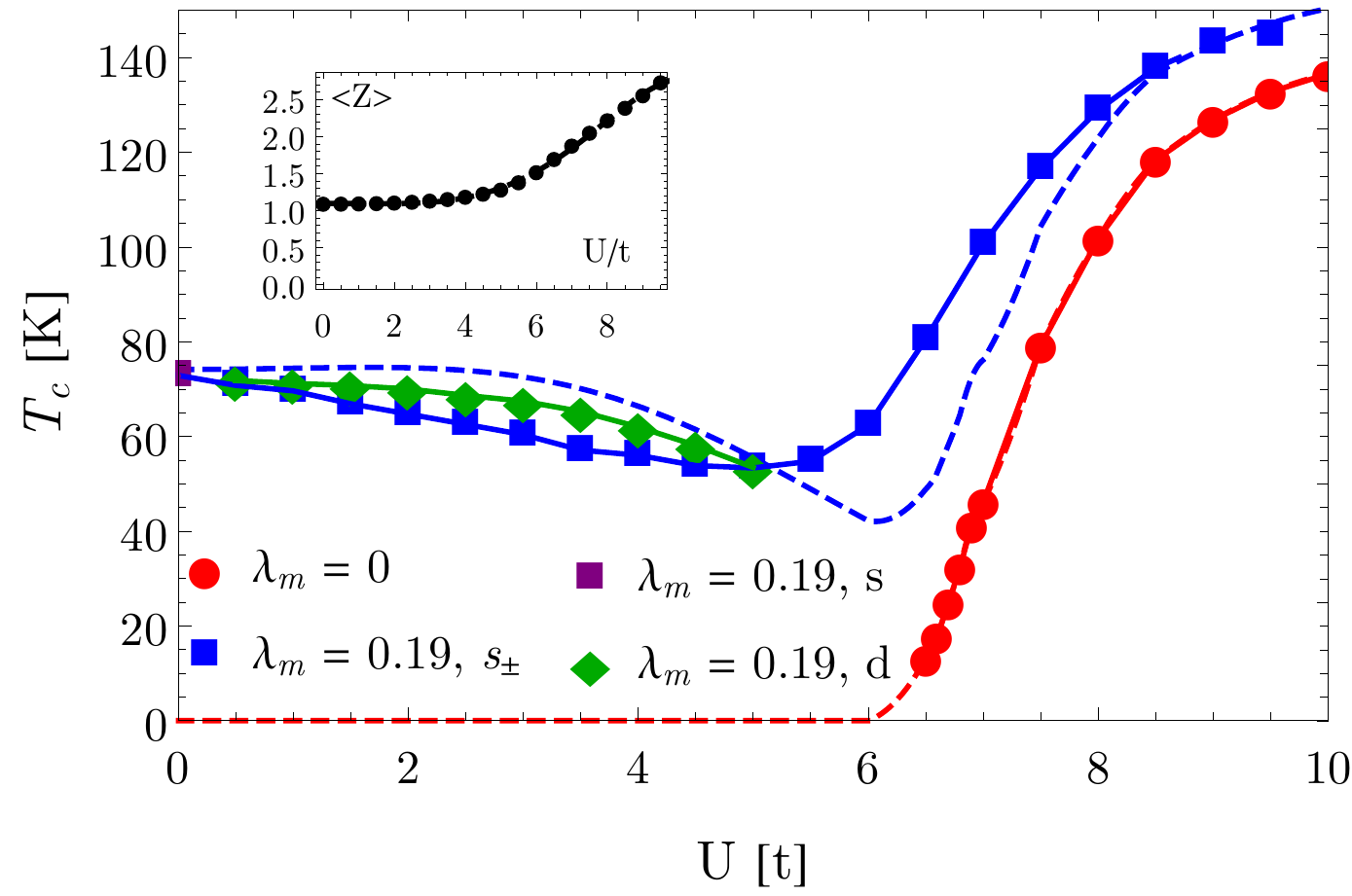}
    \caption{The superconducting critical temperature $T_c$ versus $U$ for two values of the phonon coupling $\lambda_m$. The symmetry of the order parameter in each case is indicated in the legend. 
    The dashed lines follow the phenomenological formula of Eq.~\eqref{Eq:TcvsULambda}. In the absence of electron-phonon coupling, spin fluctuation superconductivity is exponentially suppressed up to a critical value of $U\approx 6.5t$. When including electron-phonon coupling, introducing $U$ slowly decreases $T_c$ due to quasiparticle renormalization, only to increase again when $U \gtrsim 5t$. In the intermediate regime $0 < U < 5t$, a $d$-wave solution has the highest $T_c$. {\em Inset:} The Fermi-surface averaged value of $Z$, where $1/Z$ reflects the quasiparticle pole strength.}
    \label{Fig:FigTcvsU}
\end{figure}

\begin{figure}[t]
    \centering
    \includegraphics[width=\columnwidth]{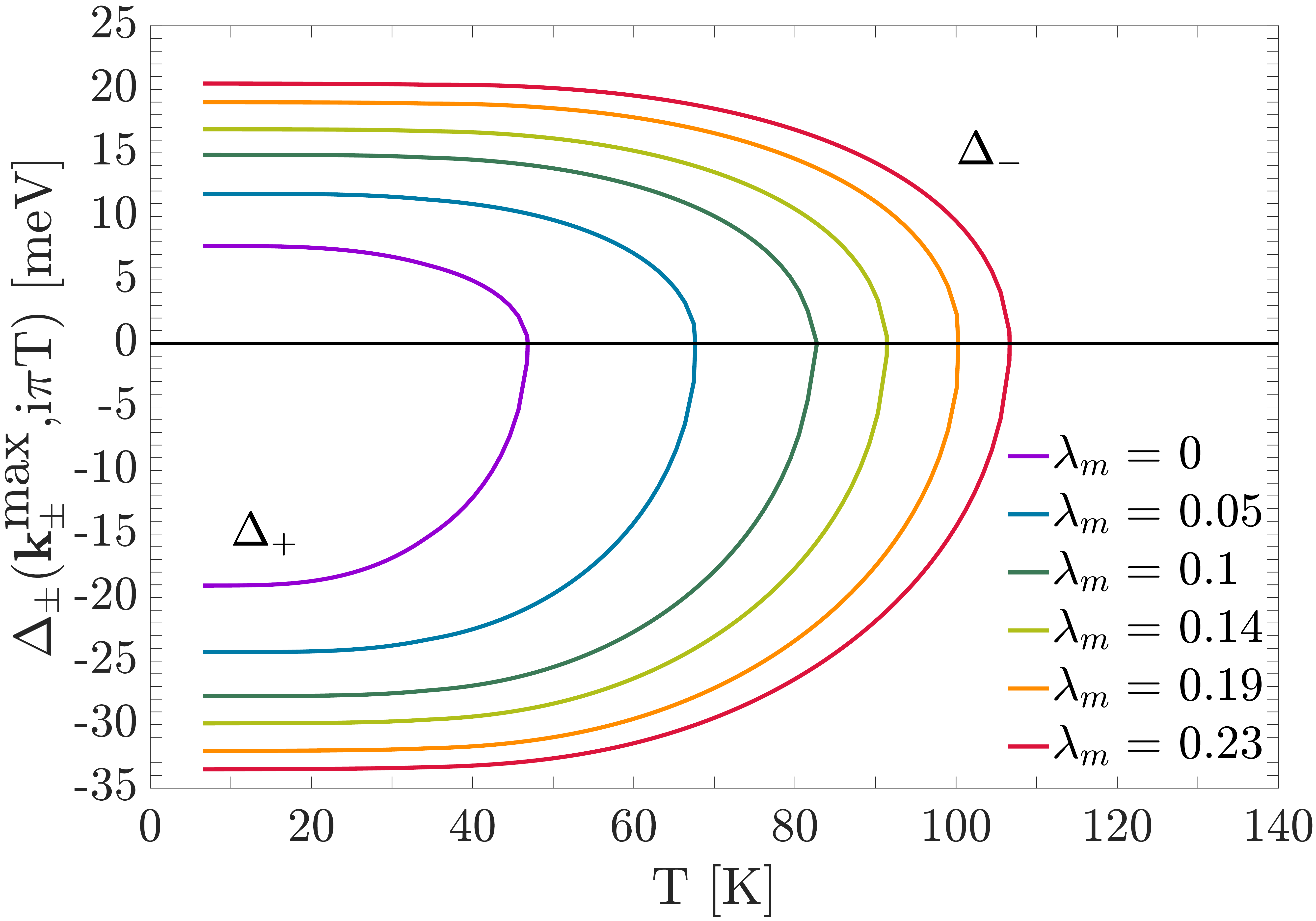}
    \caption{The maximum value of the superconducting gap on the band crossing the Fermi level ($\Delta_-$) and the incipient band ($\Delta_+$) at the smallest Matsubara frequency 
    $\Delta({\bf k}_\pm^\text{max},\text{i}\pi T)$ as a function of temperature. The solid lines show the results for $U=7t$ for various strengths of the electron-phonon coupling $\lambda_m$, as indicated in the legend. 
    The values of $t_\perp$ used here are the same as those used in Fig.~\ref{fig:FigTcvslambda}.
    }
    \label{fig:FigGapvsT}
\end{figure}

Next, we determined $T_c$ by tracking the maximum value of the anomalous self-energy $\phi_{\pm}({\bf k})$ or, equivalently, the gap function $\Delta_\pm({\bf k})$ as a function of temperature (for example, see Fig. \ref{fig:FigGapvsT}). As shown in  Fig.~\ref{Fig:FigTcvsU}, we find that $T_c (U)$ is exponentially suppressed for small $U$, only to significantly increases around $U=6.5t$ in the absence of {\eph} coupling. For a reasonable value of $U=7t$ ($=0.54$~eV for $t=75$ meV), we obtain a $T_c = 46.8$~K, which is slightly larger than the $T_c$ in surface-doped FeSe thin films. 

\subsection{Superconductivity with electron-phonon coupling}
Having established the superconducting properties of the bilayer Hubbard model predicted by FLEX, we now turn to the case with {\eph} coupling. ME theory predicts that having only forward phonon scattering ($U = 0$, $\lambda_m\ne 0$) yields an approximately linearly increasing $T_c \approx \frac{\lambda_m}{2+3\lambda_m} \Omega$ as a function of $\lambda_m$, see Fig.~\ref{fig:FigTcvslambda}~\cite{Rademaker2016}. The resulting pairing symmetry is $s$-wave but with a gap structure that is strongly peaked near the Fermi level~\cite{WangSUST}. To illustrate this limit, Fig.~\ref{fig:Bands}(b) plots the bare and renormalized band structures together with the gap function for the case $U = 0$, $t_\perp = 4.21t$, and  $\lambda_m = 0.19$. Here, we observe relatively weak band renormalizations in comparison to the case when $U \ne 0$. Note that the renormalized bands plotted in Fig.~\ref{fig:Bands} correspond to the main band, only, and do not include the replica bands. Our previous work \cite{Rademaker2016} demonstrated that this value of $\lambda_m$ also produces replica bands with a shape and total spectral weight that is consistent with experiments \cite{Lee2014}. This value of $\lambda_m$ produces a superconducting state with $T_c \approx 60$ K within ME theory.

Having established the relevant $(U \ne 0, \lambda_m = 0)$ and $(U = 0,\lambda_m \ne 0)$ limits, we now study the interplay between the spin- and {\eph} interaction 
by fixing $\lambda_m = 0.19$ and slowly increasing the value of $U$, as shown in Fig~\ref{Fig:FigTcvsU}. Beginning 
from $U = 0$, we find that the Hubbard repulsion initially leads to a {\em decrease} of $T_c$. This reduction occurs because the spin-fluctuation contribution to superconductivity is exponentially suppressed for small $U$ while the corresponding decrease in the quasiparticle pole is non-negligible. This effect is captured by expanding the normal self-energy as $\Sigma ({\bf k}, \mathrm{i} \omega_n) = \mathrm{i} \omega_n  [1 - Z({\bf k},\mathrm{i} \omega_n)] + \chi({\bf k},\mathrm{i} \omega_n)$, where $1/Z$ reflects the quasiparticle pole strength~\cite{Imada:1998er}. 
We can gain a qualitative understanding of the effect of a non-zero $Z$ by considering the perfect forward scattering limit. Perturbation theory predicts $Z = 1 + \mathcal{O}(U^2)$ in the weak coupling limit. To lowest order, we can incorporate this 
aspect into the weak coupling calculation presented in Sec. 2 of Ref.~\cite{Rademaker2016}, by replacing the assumption $Z({\bf k},\mathrm{i}\omega_n) = 1$ by $Z({\bf k},\mathrm{i}\omega_n) = Z$, a frequency and momentum-independent constant. This approximation leads to 
\begin{equation}
    T_c (\lambda_m,U) = \frac{\lambda_m\Omega}{2Z^2+3\lambda_m}, 
    \label{Eq:TcvsLambda}
\end{equation}
which is valid in the limit $U/t \ll 1$. Eq.~\eqref{Eq:TcvsLambda} indicates that the 
superconducting $T_c$ actually {\em decreases} with small $U$ due to the overall increase 
in the quasiparticle mass. A similar conclusion was also obtained in the five-orbital 
treatment of the problem presented in Ref.~\cite{Schrodi:2020bq}. 

Interestingly, we also observe several changes in the pairing symmetry of the system 
as the value of $U$ increases. For $U = 0$, the system 
has an $s$-wave gap, as shown in Fig. \ref{fig:Bands}b. 
When $0<U \le 5t$, we find that a $d$-wave solution with nodes on the Fermi surface has a higher $T_c$. Note that Ref.~\cite{Schrodi:2020bq} also found a $d$-wave solution in the same limit, albeit in their case with a nodeless $d$-wave gap. For stronger interactions $U\gtrsim 5t$, the $s_\pm$ solution takes over again and has the largest $T_c$. Note that $s_\pm$ becomes favorable because we adjust $t_\perp$ to ensure that the incipient band remains close to the Fermi level in the renormalized bandstructure. If one does not adjust $t_\perp$, the Hartree contribution will push the incipient band down in energy, and consequently the $d$-wave solution would remain favorable.

For even stronger coupling $U>6.5t$, when the spin fluctuations are strong enough to support superconductivity by themselves, we obtain a combined effect between the {\eph} coupling and the FLEX mechanism. In this region, the {\eph} coupling enhances the $T_c$, and the system has an $s_\pm$ pairing symmetry. The size of the enhancement, however, depends on the relative strength of the two pairing channels; for large $U$, the total $T_c$ begins to approach the values obtained in the absence of {\eph} interaction. 

To investigate the combined contributions of the two pairing mediators further, we now focus on a particular parameter set that reproduces the $T_c$ values in FeSe and its cousins. Specifically, we fix $U = 7t$, which results in a spin-fluctuation mediated $s_\pm$ state with a $T_c = 46$ K when $\lambda_m = 0$. We then increase $\lambda_m$ and monitor the evolution of the pairing symmetry and $T_c$. In this case, we always obtain an $s_\pm$ order parameter and a $T_c$ that increases (approximately) linearly with $\lambda_m$, as shown in Fig.~\ref{fig:FigTcvslambda}. However, we also observe that the slope of $T_c$ vs $\lambda_m$ is smaller for $U = 7t$ in comparison to the $U =  0$ case, in qualitative agreement with the behavior predicted by Eq.~\eqref{Eq:TcvsLambda}. We conclude that in the regime for $U \gtrsim 7t$ and small $\lambda_m$, the spin fluctuation pairing and the forward scattering act as parallel channels. This conclusion is in contrast to Ref.~\cite{Schrodi:2020bq}, which always obtained a $T_c$ suppression. 

We have found that $T_c$ as a function of $\lambda$ and $U$ can be approximated, in the experimentally relevant limit, using a simple phenomenological model
\begin{equation}
    T_c (U,\lambda_m) = T^{\mathrm{SF}}_c (U) + T^{\mathrm{e-ph}}_c (\lambda_m).  
    \label{Eq:TcvsULambda}
\end{equation}
Here, $T^{\mathrm{e-ph}}_c$ is obtained using Eq.~\eqref{Eq:TcvsLambda} and 
$T^{\mathrm{SF}}_c (U)$ is value of $T_c$ obtained from our FLEX 
calculations when $\lambda_m = 0$ (the red dots in Fig.~\ref{Fig:FigTcvsU}).  
When computing  $T^{\mathrm{e-ph}}_c$, we approximated $Z$ with a Fermi surface average $Z = \sum_{\alpha = \pm}\langle Z_\alpha({\bf k},\text{i}\pi T) \rangle_\mathrm{FS}$, where $Z_\pm({\bf k},\text{i}\pi T)$ is obtained from our final converged solution for $U=7t$ and $\lambda_m = 0$. The resulting $T_c (U,\lambda_m)$, plotted as the blue dashed line in Fig.~\ref{Fig:FigTcvsU}, qualitatively captures the non-monotonic evolution of the full self-consistent results.

\subsection{The $\Delta/T_c$ ratio}
A standard measure of the pairing interaction's strength is the deviation of the ratio $\Delta(T=0)/T_c$ from the BCS value of 1.764. For example, unconventional superconductors in the strong coupling limit typically have values of $\Delta(T=0)/T_c > 5$~\cite{Inosov:2011bj}.  
Fig.~\ref{fig:FigGapvsT} plots the gap at the lowest Matsubara frequency $\Delta_\pm({\bf k}^\text{max}_\pm,\text{i}\pi T) = \phi_\pm({\bf k}^\text{max}_\pm,\text{i}\pi T) / Z_\pm({\bf k}^\text{max}_\pm,\text{i}\pi T)$ vs temperature for a variety of model parameters. 
Here, ${\bf k}^\text{max}_\pm$ is the momentum where the gap function takes on its largest 
absolute value on the $+$ and $-$ bands. 

As discussed in Ref.~\cite{WangSUST}, the gap function produced by a 
forward-focused \eph~interaction only ($\lambda_m \ne 0,~U=0$) is strongly peaked on the 
Fermi surface, as shown in Fig.~\ref{fig:Bands}(b). In this case, we find that the $\Delta/T_c$ ratio 
is between 2 and 2.3, consistent with previous works ~\cite{WangSUST,Schrodi:2020bq}. 
This value is slightly larger than those obtained for weak-coupling conventional superconductors. 
The spin fluctuation mechanism alone, on the other hand, leads to larger gap values and 
a gap function that varies strongly with both momentum and band index. 
When $U \ne 0$ and $\lambda_m = 0$ [see Fig.~\ref{fig:Bands}(a)], the global maximum 
of $|\Delta_\pm({\bf k},\text{i}\pi T)|$ occurs at the $M$-point and on the incipient ($E_+({\bf k})$) band, while the largest value of the gap on the $E_-({\bf k})$ band occurs at the $\Gamma$-point.  
When the \eph~interaction is introduced, however, $\Delta_{-}({\bf k},\text{i}\pi T)$ develops a peak centered at ${\bf k}_\mathrm{F}$ such that the local maximum on the $E_-({\bf k})$ band  
shifts to the Fermi surface while the global maximum remains at the $M$ point [see Fig.~\ref{fig:Bands}(c)]. 

\begin{figure}[t]
    \centering
    \includegraphics[width=\columnwidth]{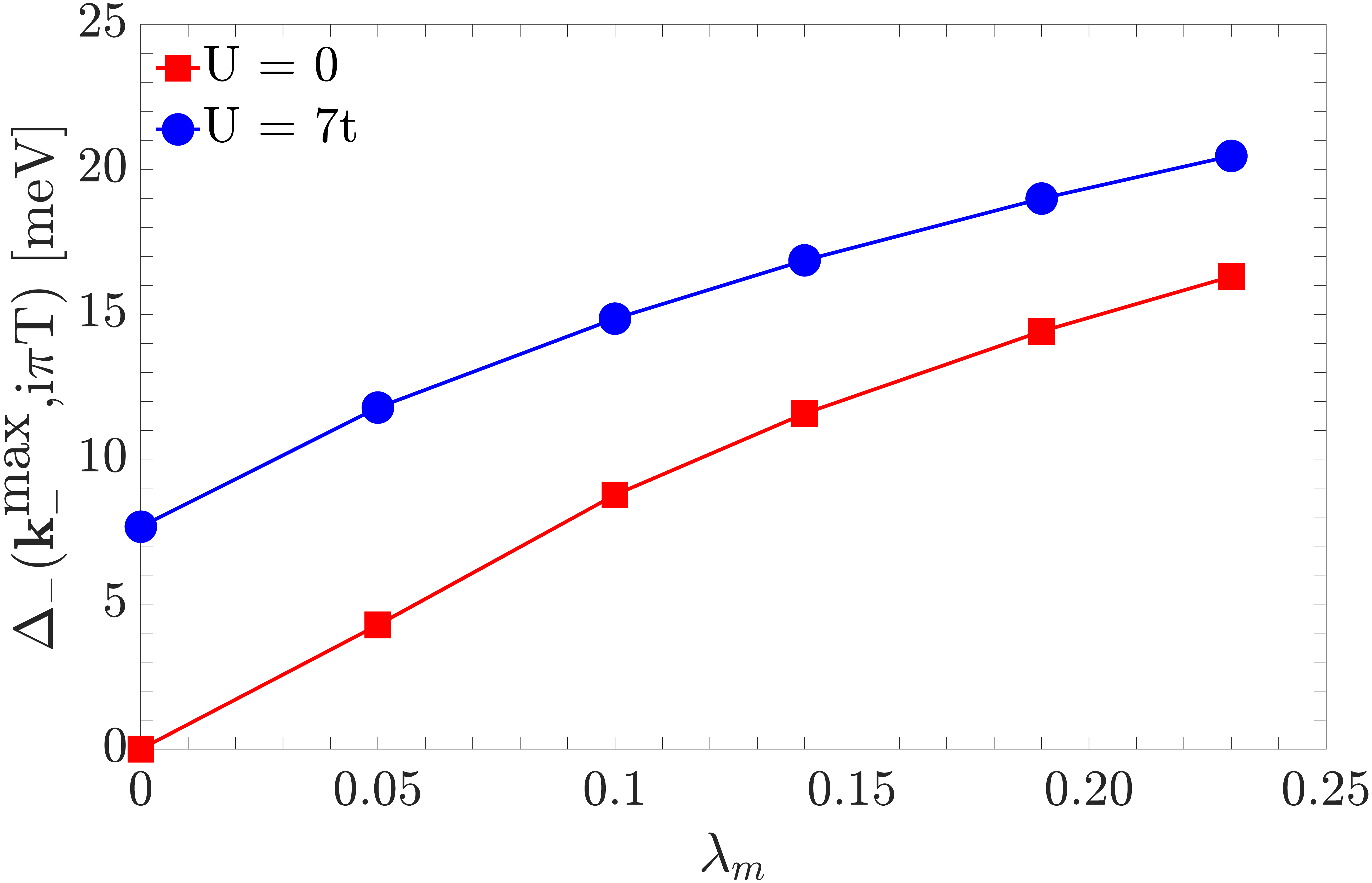}
    \caption{The maximum value of the superconducting gap on the Fermi surface of the antibonding band 
    $\text{max}|\Delta_-({\bf k}_\text{F},\text{i}\pi T)|$ as a function 
    of $\lambda_m$ for the same parameters used in Fig.~\ref{fig:FigTcvslambda}. 
    Here, the location of the gap maximum corresponds to a momentum point on the 
    Fermi surface ${\bf k}^\text{max}_- = {\bf k}_\mathrm{F}$ and, therefore, 
    approximates the spectral gap. 
    } 
    \label{fig:FigDeltavslambda}
\end{figure}

To study the $\Delta/T_c$ ratio in the combined spin-fluctuation plus \eph~ scenario, Fig.~\ref{fig:FigGapvsT} plots the temperature dependence of the maximum value of the 
gap function on both bands for $U = 7t$ and several values of $\lambda_m$. With the exception of 
the $\lambda_m = 0$ case, ${\bf k}_-^\text{max} = {\bf k}_\mathrm{F}$ for all values of 
$\lambda_m$ while ${\bf k}_+^\text{max} = (\pi/a,\pi/a)$ in each case. 
Since the $\Delta/T_c$ ratio in FeSe/STO has typically been determined using spectroscopic techniques, it is the maximum value on the Fermi surface is the relevant. Fig.~\ref{fig:FigDeltavslambda}, 
therefore, summarizes the evolution of $\Delta_{-}({\bf k},\text{i}\pi T)$ as a function 
of $\lambda_m$, where we find that the forward-focused \eph~interaction increases the 
gap at $E_\text{F}$, consistent with the behavior of $T_c$ shown in Fig.~\ref{fig:FigTcvslambda}. 
In this case, we obtain the $\Delta / T_c = 1.90$ for $\lambda_m = 0$, which is enhanced 
slightly to $\Delta / T_c = 2.23$ for $\lambda_m = 0.23$. These values are reduced slightly from the value obtained with only the \eph~interaction and they are comparable to the experimental value $\Delta/T_c\approx 2.12$ in FeSe/STO~\cite{Song:2019cg,Lee2014}. 

Since the ratio of the intensity of the replica and main bands 
$\eta = I_1/I_0 \propto \lambda_m$ \cite{Rademaker2016}, our results also 
qualitatively explain the observed correlation between the spectroscopic 
superconducting gap and $\eta$ \cite{Song:2019cg}. We note, however, that Ref.~\cite{Song:2019cg} 
observed a slower variation with $\Delta(\eta) \approx \Delta_0 + \Delta_1 \eta/2$, whereas our model 
predicts $\Delta(\eta) \approx \Delta_0 + \Delta_1 \eta$. Increasing the value of 
$q_0$ to reduce the forward-focus of the \eph~interaction might resolve this issue, however, more 
detailed comparisons between the predicted and measured spectral functions are needed to address 
this issue. We also cannot exclude the possibility that a more exact treatment of the electron-electron interaction would produce a smaller superconducting gap. There is, in fact, some precedent for this as FLEX consistently over predicts the gap magnitude in the cuprates, see Ref.~\cite{Monthoux:1994jn}. 
Nevertheless, our results demonstrate that the combined pairing scenario is not only compatible 
with an incipient $s_\pm$-pairing scenario, but that it can also account for the qualitative $T_c$ and 
$\Delta$ enhancements observed experimentally. 

\section{Summary and Conclusions}
\label{Sec:Outlook}
We have shown that for small $\lambda_m$ and large Hubbard $U$ values, the spin fluctuation and forward-focused electron-phonon mechanisms for superconductivity can work in unison, providing a significant enhancement of $T_c$ in the $s_\pm$ pairing channel. Moreover, the resulting values of the total $T_c$ and $\Delta/T_c$ ratios are consistent with experiments for monolayer FeSe/STO. 

Recently, Ref.~\cite{Schrodi:2020bq} reported a combined FLEX plus forward scattering calculation within a five-band model of FeSe/STO (see also Ref.~\cite{Schrodi:2020bz}). That work found that the combined pairing mediators produced a nodeless $d$-wave gap symmetry, and a complicated dance of collaboration and competition between interfacial phonons and the spin fluctuations. We have obtained similar results using a two-orbital model when $U \lesssim 5t$, which suggests that the physics in this region of parameter space is robust against variations in the details of the model. The most notable difference between Ref.~\cite{Schrodi:2020bq} and our work is that we self-consistently calculate the spin excitation spectrum, rather than using spin susceptibility computed using the non-interacting band structure. This difference is an important one for several reasons. First, we showed that the Hartree contribution to the self-energy produces a relative shift of the electron- and hole-like bands, leading to a qualitative change in the band structure before and after self-consistency, which in turn affects the competition between $d$-wave and $s_\pm$-wave solutions. Second, the self-consistent calculation of the interaction kernel allows us to go to further into the strong coupling limit for the Hubbard interaction, which is otherwise forbidden by the Stoner criterion. Ideally, further study of a fully self-consistent (and numerically challenging) FLEX calculation for a five-band model is necessary to quantify the difference between both methods.

Regardless of the differences, the most critical test of a theory is experimental results. In principle, a difference between nodeless $d$-wave and $s_\pm$-wave superconductivity can be established using phase-sensitive measurements. Further study might elucidate the effects of impurities on both these states, which can be probed using scanning tunneling microscopy~\cite{Fan:2015dl, Zhang:2020fg}.  Furthermore, the gap-to-$T_c$ ratio can indicate the relative importance of either spin fluctuations or phonon coupling. Note, however, that there is some ambiguity in establishing the experimental value of the critical temperature, as the gap opening in spectroscopic measurements doesn't necessarily happen at the same temperature as the onset of the resistance transition in transport measurements~\cite{Faeth:2020un}. This question cannot be answered within the framework of Eliashberg and requires the inclusion of the BKT mechanism into the theory. Despite these theoretical subtleties, our results do indicate that $s_\pm$ superconductivity is, in principle, possible when spin fluctuations and electron-phonon forward scattering collaborate.

\begin{acknowledgments}
The authors thank T. A. Maier for useful discussions and comments on the manuscript. This work was supported by the Office of Naval Research under Grant No. N00014-18-1-2675. L.~R. was supported by an Ambizione grant from the Swiss National Science Foundation.
\end{acknowledgments}

\bibliography{references}% Produces the bibliography via BibTeX.

\end{document}